\documentclass[namedreferences]{solarphysics}
\usepackage[natbib,optionalrh,solaromanenum]{spr-sola-addons} 
\usepackage{graphicx}                    
\usepackage{amssymb}                    
\usepackage{color}                       
\usepackage{url}                         

\begin{document}

\begin{article}

\begin{opening}

\title{Photon mean free paths, scattering, and ever-increasing telescope resolution}

%
\author{P.G~\surname{Judge}$^{1}$\sep
        L.~\surname{Kleint}$^{2}$\sep
        H.~\surname{Uitenbroek}$^{3}$\sep
        M.~\surname{Rempel}$^{1}$\sep
        Y.~\surname{Suematsu}$^{4}$\sep
        S.~\surname{Tsuneta}$^{4}$\sep
       }

%
\runningauthor{Judge et al. }
\runningtitle{photons, scattering, spatial resolution}

%
  \institute{$^{1}$ High Altitude Observatory, National Center for Atmospheric Research$^5$, email: \url{judge@ucar.edu, rempel@ucar.edu} \\ 
             $^{2}$  Institute of 4D Technologies, University of Applied Sciences and Arts Northwestern Switzerland, 5210 Windisch, Switzerland \url{lucia.kleint@fnhw.ch}\\
             $^{3}$ National Solar Observatory, email: \url{huitenbroek@nso.edu} \\
             $^{4}$ National Astronomical Observatory of Japan, email: \url{suematsu@solar.mtk.nao.ac.jp, saku.tsuneta@nao.ac.jp} \\
$^5$ {The National Center for Atmospheric Research is sponsored by the National Science Foundation}
             }

\begin{abstract}
We revisit an old question: what are the effects of observing
stratifed atmospheres on scales below a photon mean free path $\lambda$?
The mean free path of photons {\em emerging} from the
solar photosphere and chromosphere is $\approx 10^2$ km.  
Using current 1m-class telescopes, $\lambda$ is on the
order of the angular resolution.  But the 
Daniel K. Inoue Solar Telescope will have a diffraction limit of
$0.020''$ near the atmospheric cutoff at 310nm, corresponding to 14 km at the solar surface.
Even a
small amount of scattering in the source function leads to physical
smearing due to this solar ``fog'', with effects similar to a
degradation of the telescope PSF.  
We discuss a unified picture that depends simply on the nature and 
amount of scattering
in the source function.   Scalings 
are derived from which 
the scattering in the solar atmosphere can be 
transcribed into an effective  Strehl ratio, a quantitity useful to observers. 
Observations in both permitted ({\em e.g.}, 
{Fe~{\sc i}} 630.2 nm) and forbidden ({Fe~{\sc i}} 525.0 nm) lines 
will shed light on both instrumental performance as
well as on small-scale structures in the solar atmosphere.
\end{abstract}

%

\end{opening}

%


\newcommand{\be}{\begin{equation}}
\newcommand{\ee}{\end{equation}}
\newcommand{\pref}[1]{\protect\ref{#1}}

\newcommand{\tabone}{
%
 \begin{table}
\label{tab:lines}
 \caption{Spectral lines commonly used to determine thermal and magnetic structure}
 \begin{tabular}{lllllll}     
 \hline
$\lambda$  nm   &log gf&   Ion & Lower level &  Upper level & $z$ km& $-\log_{10} \epsilon$\\
 \hline
422.6728 &   +0.243 &  {Ca~{\sc i}} & $4s^2~^1\!S_0 $& $4s4p~^1\!P^o_1$ & 200-1000$^a$ & 2.9--4.5$^h$    \\
524.7049 &   -4.946 &  {Fe~{\sc i}}&  $a~^5\!D_2$ &    $z~^7\!D^o_3$ & 0--500$^b$ & 0--0.01\\
525.0208 &   -4.938 &  {Fe~{\sc i}}&  $a~^5\!D_0$ & $z~^7\!D^o_1$ & 0--500$^b$ & 0--0.02\\
589.5924 &   -0.184 &  {Na~{\sc i}}  &  $3s~^2S_{1/2}$ &  $3p~^2P^o_{1/2}$  &50--700$^c$& 1.4--3.7\\
617.3341 &   -2.880 &   {Fe~{\sc i}}&  $a~^5\!P_1$ & $y~^5\!D^o_0$ &  20--300$^d$ & 1.0--2.3\\
630.1498 &   -0.745 &   {Fe~{\sc i}}&  $z~^5\!P^o_2$ & $e~^5\!D_2$ &  20--210$^e$&0.9--1.9\\
630.2494 &   -1.203 &   {Fe~{\sc i}}&  $z~^5\!P^o_1$ & $e~^5\!D_0$ &  20--210$^e$&0.9--1.9\\
676.7768  &  -2.170 & {Ni~{\sc i}} & $a~^1\!S_0$ & $z~^3\!P^o_1$   & 250-450$^f$ &  0.5--1.0\\
709.0378 &   -1.210 &   {Fe~{\sc i}}&  $y~^5\!D^o_1$ & $e~^5\!F_1$ & 200-500$^a$ & 1.7--2.8\\    
854.2091 &   -0.362 &   {Ca~{\sc ii}}&      $3d~^2\!D_{5/2}$  & $4p~^2P^o_{3/2}$ &200--1300$^g$&0.8--2.6\\
\hline
\end{tabular}\\
Notes: Oscillator strengths are from 1995 Atomic Line Data
(R.L. Kurucz and B. Bell) Kurucz CD-ROM No. 23. Cambridge, Mass.:
Smithsonian Astrophysical Observatory.  Values of $\epsilon$ are computed 
for the range of formation heights listed, 
using the
Van Regemorter approximation for permitted transitions. For the spin forbidden
transitions a cross section of $\pi a_0^2$ was used where $a_0$ is the Bohr radius. 
Formation heights are from: 
$a$ from VAL model 3C, computed in this work;
$b$ 525.0 from \cite{Lites1973}, with 325 km added (the limb-to-center viewing angle
correction, {\em e.g.\rm} \cite{Athay1976}), 524.7 
assumed the same; 
$c$ \cite{Leenaarts+others2010}; 
$d$ \cite{Norton+others2006}; 
$e$ \cite{Faurobert+others2013}; $f$
\cite{Bruls1993};  $g$ \cite{Cauzzi+others2008}. $h$ A collision strength of 14.4 was
adopted here instead of the Van Regemorter approximation. 
\end{table}
}

\newcommand{\figone}{
\begin{figure}
\label{fig:s} 
\includegraphics[width=12.0cm]{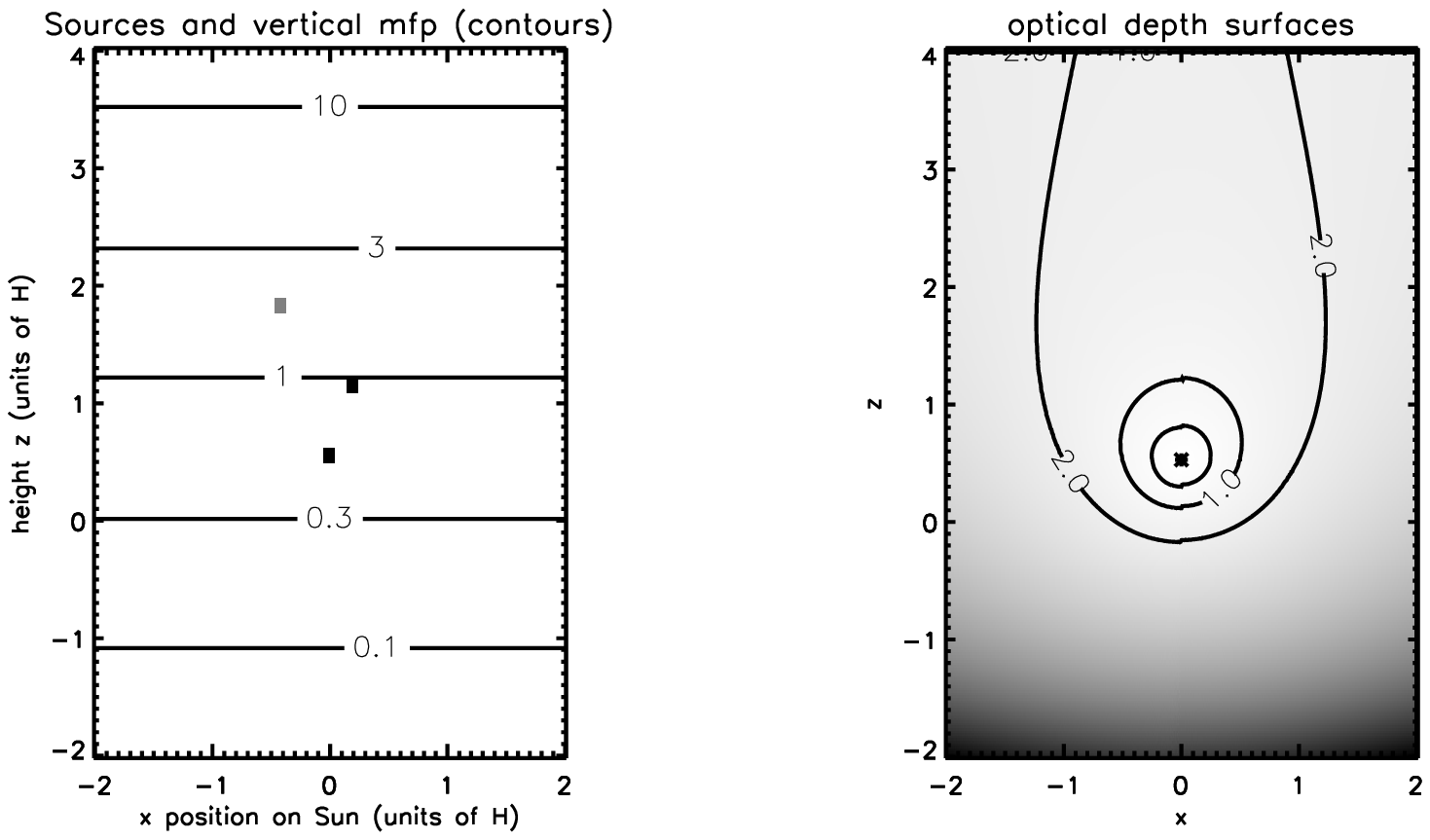}  
\caption{Left:  Isolated sources of radiation embedded in a
  stratified, scattering atmosphere are shown.  No thermal sources of emission other 
than the three bright points shown were included. 
Contours show the local mean free
  path
length.  All distances are in units of the local density scale height
$H$.  The two sources are placed just below the $\tau=1$ surface
(contour of mfp =1) and are separated by 1/2 of a pressure scale
height.
Right: surfaces of constant optical depth are shown
as an image (reverse colors, dark=larger values) and with contours, 
calculated from
the deepest of the sources shown in the left hand panel.  The $\tau=1$ surface is egg-shaped. Radiation escapes
preferentially in the outward (positive $z$) direction.  
}
\end{figure}
}

\newcommand{\figthree}{
\begin{figure}
\label{fig:compare} 
\includegraphics[width=8cm]{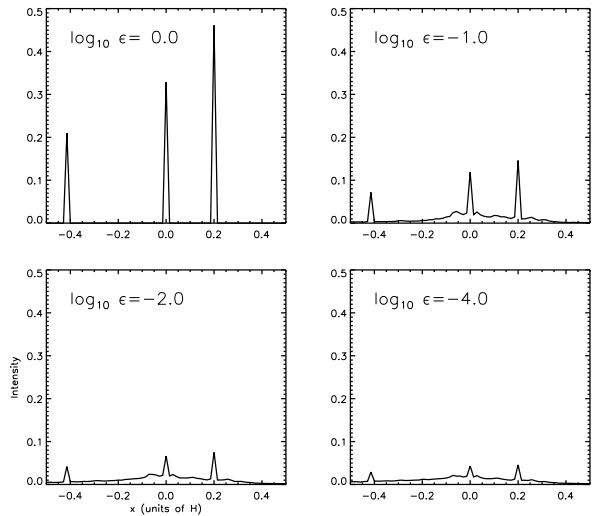}
\caption{Emergent intensity as a function of $x$ as
  seen from an inclination of zero.   The integrals under the curves
  are the same. 
}
\end{figure}
}

\newcommand{\figvr}{
\begin{figure} 
\label{fig:vanreg} 
\includegraphics[width=10cm]{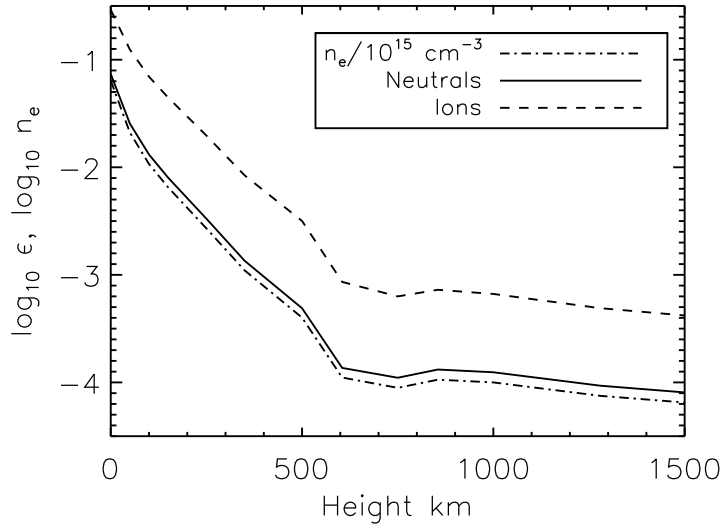}
\caption{Destruction probabilities are shown
as a function of height for the model 'C' atmosphere of
\cite{Vernazza+Avrett+Loeser1981}, for permitted transitions, 
using Van Regemorter's (1962) approximation.  A wavelength of 500 nm
was used to make this plot.
}
\end{figure}
}

\newcommand{\figthreed}{
\begin{figure}
\label{fig:threed} 
\includegraphics[width=11.5cm]{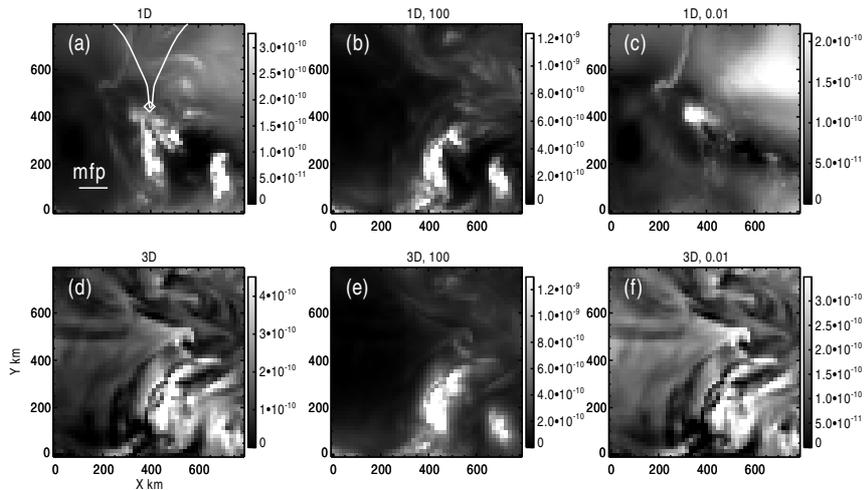}
\caption{Images in the core of the {Ca~{\sc i}}
resonance line computed from a 3D magnetoconvection model.   
The top row shows 
intensities computed from
1D transfer calculations along the vertical for each individual
pixel.   The left panel also show the wavelength marked 
with a diamond symbol plotted over the mean line profile.
The second row shows the full 3D calculations.
From left to right, the values of $\epsilon$ used were 1, 100 and 0.01 respectively.
The images correspond to about $1\times1$ seconds of arc,
as observed from Earth, 
and the ``pixels'' are 16 km or about 1/45 of an arcsecond.
}
\end{figure}
}

\newcommand{\figscales}{
\begin{figure} 
\label{fig:scales} 
\includegraphics[width=10cm]{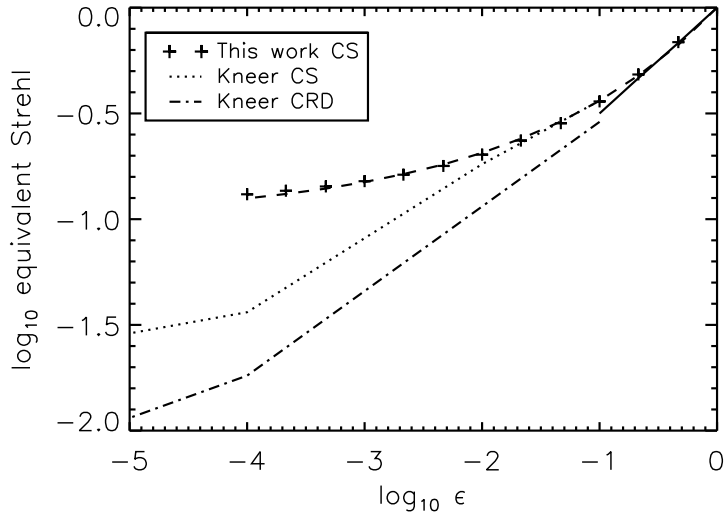}
\caption{The peak brightness, normalized to the case of no
scattering ($\epsilon=1$) taken from the coherent scattering 
calculation data shown in Figure~2.  The ordinate is one useful definition of the equivalent Strehl ratio. 
The short solid line is ${\cal S} = \sqrt{\epsilon}$, the dashed line $\log_{10}{\cal S} = \epsilon^{0.26}-1$.
 }
\end{figure}
}

\def\jnu{\ifmmode{j}_{\nu}\else${j}_{\nu}$\fi}
\def\anu{\ifmmode{\alpha}_{\nu}\else${\alpha}_{\nu}$\fi}
\def\uniti{erg~$\,$cm$^{-2}\,$\-s$^{-1}\,$\-hz$^{-1}\,$\-sr$^{-1}$}
\def\unite{erg~$\,$cm$^{-3}\,$\-s$^{-1}\,$\-hz$^{-1}\,$\-sr$^{-1}$}
\def\inm{\ifmmode{I}_{\nu\mu}\else${I}_{\nu\mu}$\fi}

\section{Introduction}\label{s:Introduction} 

The photosphere and lower chromosphere are hydrostatically stratified
atmospheric layers of the Sun from which the bulk of the solar
radiation emerges.  
In this paper we study the effects of observing
these regions on scales below $\lambda$, the mean free path (mfp) of photons. 
Our goals are to explain some physical effects resulting from photon
scattering, to examine basic limitations that the Sun itself presents,
and to quantify their dependence on the nature  of the scattering.
On scales currently observable with 1m class telescopes near 500nm,
$\lambda $ is on the order of the angular resolution.  But 
the 4m 
Daniel K. Inoue Solar Telescope will
have a diffraction limit of $0.020''$ near 310nm, $0.038''$ near
600nm, corresponding to 14 and 27 km at the solar surface.  These are
mere fractions of $\lambda$.

In the Sun's atmosphere, magnetoconvection and other processes 
can generate thermal
and magnetic structure on scales significantly below $\lambda$.
This is because local radiative losses and gains do not necessarily occur over
spatial scales dictated by the continuum mean free path. The gains/losses 
arise from integrals over the whole spectrum, including many lines far more opaque
than the neighboring continuum, with far smaller photon mfps.   MHD
processes create structures with non-zero temperature contrast on
spatial scales much smaller than the continuum mean free
path \citep[{\em e.g. \rm}][]{Knoelker+others1988}. 
We do not address this problem here. Instead we examine 
limitations imposed by photon scattering.

Many authors have explored this problem using 
spectral lines with diverse scattering properties. We 
will show that the various calculations in the literature 
are readily understood taking into consideration the ``scattering parameter''
$\epsilon$. Defined below (equation~\pref{eq:parameter}), this measures the 
probability of photon destruction per interaction with the atom.
A list of spectral lines commonly used to explore
the Sun's magnetic and thermal atmospheric structure is given in Table~1. Note that they
too have a variety of scattering parameters. 

\tabone

``Spatial resolution'' is a problem not only
in the usual, ``plane-of-the-sky'' sense, but also along 
the line-of sight (LOS) direction.  This is  of physical  interest, because
measurements of LOS derivatives in the magnetic field vector 
{\bf B}, in particular, are important for physical and technical reasons
\citep[e.g.\rm][]{Socas-Navarro2005a}.
The ability of a telescope/\-spectrograph to resolve features along the 
LOS has been discussed by Landi degl'Innocenti (2013)\nocite{Landi2013}. 
Within a stratified atmosphere the 
thickness of the spectrum\--forming layer -- that from which the photons last interact with the matter 
-- is always roughly a local
pressure scale height.  We show this formally in Appendix \pref{sec:emerging}, 
and review why the intensity regulated by the H$^-$ ion, 
which dominates the visible continuum opacity, forms naturally over half the distance.
To look at the ability of telescopes to resolve features in the
  POS, below we examine radiative transfer calculations including varying
amounts of photon scattering. To understand the calculations we use equations 
for the source function $S$ of the form
\be \label{eq:parameter}
S = (1 - \epsilon) J + \epsilon B.
\ee
Here $B$ is the Planck function and 
the meanings of $\epsilon$ and mean intensity $J$ are different for
coherent scattering and for incoherent scattering in spectral lines (see below).

The development of efficient transfer algorithms in the 1980s 
\citep[{\em e.g.\rm}][]{Scharmer+Carlsson1985,Rybicki+Hummer1991} has made
multi-dimensional transfer calculations more common, particularly 
using algorithms making use of local coupling between radiation and matter.
But the literature has
tended to focus on dynamical simulations and/or effects on Stokes profiles \citep[{\em e.g.\rm}][]
{Kiselman+Nordlund1995,Leenaarts+others2010,Leenaarts+others2012,
Holzreuter+Solanki2012,Holzreuter+Solanki2013}.
We focus instead on the non- local thermodynamic equilibrium (nLTE) intensity problem alone, 
emphasizing the understanding of 
previous work and offering suggestions for spectral line observations 
with new large telescopes.
In Section \pref{sec:early} we review earlier work
addressing effects on solar images, i.e. effects of scattering in the POS. Section~\pref{sec:main}
presents radiative transfer calculations and derives scalings of 
properties of the emergent intensity on the nature of the scattering 
and on atomic parameters.  Section~\pref{sec:discussion} combines 
the results and explains some observations of very fine structures seen 
against the solar disk (filaments, spicules). 
In Appendix B we remind readers that the bulk of the chromosphere is 
close to a hydrostatic stratification
and therefore subject to the main arguments in this paper. 
(Dynamic phenomena such as spicules have attracted much recent attention,
but these must arise from far more tenuous structures, merely being at chromospheric
temperatures).

\section{Earlier work}
\label{sec:early}

\citet{Avrett+Loeser1971} presented two component
nLTE calculations in regimes close to solar.  
They solved for source functions at two points inside and
between thermally different hexagonal columns in static but stratified
layers, for a 2-level atomic model of the chromospheric {Ca~{\sc ii}} K line.
Being a two level calculation the problem is linear (source function
is linear in intensity):
\be
S_L = (1 - \epsilon) \overline{J} + \epsilon B.
\ee
where $\overline{J}= \int J_\nu \phi_\nu \, d\nu$ is the mean
intensity weighted across the line profile $\phi_\nu$ 
and $\int \phi_\nu d\nu =1$, see \cite{Mihalas1978}.
They assumed complete redistribution (CRD) in the scattering 
terms in the source functions.   They
adopted realistically small but fixed values of $2\times10^{-4}$ for
the photon scattering parameter, which we define as the destruction probability:
\begin{equation}
\epsilon = {\cal C}_{ji}/({\cal C}_{ji}+A_{ji}),
\end{equation}
where ${\cal C}_{ji}$ (s$^{-1}$) is the transition probability between
upper level $j$ and lower level $i$ induced by collisions, and
$A_{ji}$ is the spontaneous radiative transition probability.\footnote{Since in this paper we 
focus on visible and infrared wavelengths we ignore
photon destruction by continuous absorption which dominates at UV and 
mm/radio wavelengths.}  
Horizontal transport of radiation showed two essential effects. Firstly, 
for a homogeneous opacity scale but with the Planck function differing
by a factor between 2 and 3 between the components, horizontal
variations in intensity become significantly reduced (factor 2 or so)
when temperature inhomogeneities are smaller than about four times the
opacity scale height, occurring mostly in the centrally reversed
portions of the line profile (line center optical depths below 100 or
so).  Secondly, for a horizontally homogeneous Planck function
but with opacities differing by factors of four, the 2D transfer
effects cause significant deviations in emergent intensity extending
far deeper (line center optical depths of $10^ {3-4}$).

\citet{Stenholm+Stenflo1977, Stenholm+Stenflo1978} made 
2D nLTE transfer calculations in flux tube atmospheres, the latter
paper including Zeeman-induced polarization. However, they
used the spin-changing transition of {Fe~{\sc i}} 
at 525 nm which, with an oscillator strength near $10^{-5}$
has $\epsilon$ close to 1, 
minimizing effects of scattering when compared with permitted transitions
(see Table~1).

\citet{Kneer1981} reviewed 2D radiative transfer
calculations in stratified atmospheres, pointing to  
some that appeared to contradict each other. 
\citet{Kneer1981} therefore 
made nLTE linear transfer calculations using two limits- pure
coherent scattering (CS) whose diffusive behavior mimics the partially
CS regime applicable to the wings of strong lines - and complete redistribution for a 2 level
atom.  He studied harmonic variations in the Planck function in a stratified atmosphere with
opacity a function of height,
for fixed values of $\epsilon$ between $10^{-6}$ to $10^{-1}$.  His results 
confirmed that lateral
transfer, especially in the chromosphere, 
decreases the contrast in images as nLTE effects increase (decreasing $\epsilon$), 
even on scales above the resolution of telescopes then available.  Kneer introduced 
the modulation transfer function (MTF) of the calculated intensities which is readily
combined with the point spread functions of telescopes, to arrive at a quantitative 
way to study the relative smearing of solar thermal structure by solar effects (scattering) and 
by the observing instrument.    
Kneer confirmed 
and extended to CRD cases random walk arguments of 
\citet{Owocki+Auer1980}  in which the length scales 
of lateral 
transport, defined through the PSF full width at half maximum, vary as 
\begin{equation} \label{eq:scaledep}
\ell \approx a + \ln~\tau_{th}.
\end{equation}
Here, $a \approx 1$ and $\tau_{th}$, the thermalization length measured in
units of monochromatic or line center optical depth, varies asymptotically 
as $\approx \epsilon^{-1/2}$ (CS) or $\approx \epsilon^{-1}$ (CRD).  The different 
powers of $\epsilon$ arise because 
CRD over-estimates the production of wing photons from the core, enhancing 
photon escape over the CS approximation (see {\em e.g.} \citealp{Mihalas1978}).  
In effect, Kneer provided a unified picture of various calculations available in 
1980.  

The problem was revisited by \citet{Bruls+vonderLuhe2001}.  Their goal
was to show that it is possible to resolve structures on scales
significantly {\em below} $\lambda$, in the POS.  As a measure of
``resolution'' they simply looked for visible structure in emergent
intensities around two magnetic flux sheets, without discussing MTFs
or PSFs.  Their calculations of the spin forbidden lines of {Fe~{\sc
    i}} (524.7 and 525.0 nm) and the {Ca~{\sc ii}} line intensities
presents however a mixed set of conditions.  On the one hand they emphasize
the nLTE process of photoionization in which iron is over-ionized by
UV radiation. In essence, all the nLTE for the iron calculation
is in the line opacities but not, for these spin forbiddent lines,
the source function. 
 For the permitted and strong calcium lines, the
very opposite is true, and for these lines their work 
appears consistent with the earlier work of, for instance, Avrett \&
Loeser and Owocki \& Auer.

Below we will study in isolation one source of ``nLTE'' conditions- the 
scattering parameter, and we will relate the 
calculations and their parameters to an ``effective PSF'' as seen with
a perfect, diffraction-free telescope. 
Over-ionization of elements by transfer of UV radiation is arguably less
important as a smearing agent than scattering in spectral lines
observable from the ground, because UV opacities are orders of magnitude
higher than at visible wavelengths, with correspondingly smaller values of $\lambda$.

\section{Telescope resolution vs. spectrum formation}
\label{sec:main}

\subsection{Isolated thermal sources: line of sight averaging}

All spectra are formed over finite depths in a stratified atmosphere. Appendix A
shows why this depth is an opacity scale height, equal to a density scale height 
$H$ for most atomic transitions and 1/2 of this for the  H$^-$ ion. 
Let $\Delta\approx H$ describe the {\em width} of the
distribution of heights from which the photons emerge along a given
line of sight.  The centroid of this 
distribution can be determined far more accurately- all that is
required is that we have enough photons $N$ in our measurement to make
photon noise variations $\approx \sqrt N$ much smaller than $N$
itself.   Imagine for simplicity isolated 
bright point thermal sources of emission, with 
\be
S=B
\ee
in a stratified atmosphere that can also absorb photons (scattering is
dealt with below).    The configuration is shown in
Figure~1. 

\figone

If we could observe from several angles ({\em i.e.\rm} stereoscopically) 
we would be able to determine the mean formation heights
to within measurement errors provided that we capture enough photons. 
{\em But } from just one line of sight, the sources have an intrinsic 
uncertainty in their position of origin of about $\lambda \approx H \approx
10^2$ km \citep{Landi2013}. 

The condition that we have a large number $N$ of photons is nothing
more than saying that the standard deterministic transfer equation can
be applied.  At the resolution limit of telescopes, with a Planck
function near 6000K, a spectral resolution of 10m\AA, $N\approx
2\times10^7 t \ E$, where $t$ is the integration time in s, and
$E < 1$ is the transmission efficiency of the optical system.  There
are plenty of photons for reasonable values of $t$ and $E$.  Thus the {\em
  relative} centroids of the ``heights of formation'' of various lines
can be determined to most desired levels of accuracy, but their
absolute (heliocentric) coordinates have an intrinsic uncertainty of
around $H$.  The measurement becomes photon starved when trying to measure 
weak polarization signals \citep{Landi2013}. 

\subsection{Isolated thermal sources: plane of sky smearing}

In the plane of the sky, strict thermal sources can be resolved
down to a telescope resolution scale. This is because along each line
of sight, the source functions are purely local and independent of
neighboring
lines of sight.  A feature in the atmosphere is either 
intercepted by a ray or it is
not.  For pure thermal emission two sources separated by $< \lambda$
can be resolved.   The same is true for isolated optically thin 
sources in the corona (see the Discussion). 

Returning to stratified atmospheres, it must be kept in
mind that there is always a line of sight integration of scale length 
$\Delta \approx H$, so any fine scale source structure along a length $H$ (such as 
associated with current sheets surrounding magnetic flux
concentrations) is averaged along the LOS. 

\subsection{Isolated sources including coherent scattering}

Consider the more general case where the sources include 
coherently scattered photons
\be \label{eq:source}
S_\nu=\epsilon B_\nu + (1-\epsilon)  J_\nu,
\ee
where the probability of photon destruction is $ \epsilon= 
\frac{\kappa_a}{\kappa_a +\kappa_s}$, and
$\kappa_a$ and $\kappa_s$ are the opacities for pure absorption
and scattering respectively (the total opacity being 
$\kappa=\kappa_a+\kappa_s$).  The quantity $\eta =   1-\epsilon$
is the scattering albedo, and $J_\nu$ is the
angle--averaged
intensity at frequency $\nu$.  Now the formal solution to the
transfer equation is written operationally as 
\be \label{eq:lambda}
J_\nu = \Lambda [S_\nu]
\ee
where $\Lambda$ is an integral operator over all space.  The 
source function (equation~\pref{eq:source}) is 
now {\em non-local}, requiring one to solve a familiar integro-differential equation.  For simplicity, first
consider a 1D stratified atmosphere.  The range of $\Lambda$ as
measured in photon mean free paths is the ``thermalization length''
$\tau_{th}$: in essence all regions of the atmosphere that lie within
$\tau_{th}$ mean free paths of a given point influence the value of
$J_\nu$ and hence $S_\nu$ there.  Given that the vertical emergent
intensity $\approx S_\nu(\tau=1)$, we must therefore be aware that
equation (\pref{eq:lambda}) represents a {\em physical smearing} of
the original thermal source of photons $\epsilon B_\nu$.

\noindent As already noted, the value of $\tau_{th}$ in 1D atmospheres scales with 
$\epsilon$ 
as \citep{Mihalas1978}
\be \label{eqn:therm}
\tau_{th} \approx 1/ \sqrt{\epsilon} \ \ \
{\rm (coherent\ scattering),\  and } \approx 1/\epsilon \ \ \
{\rm (spectral\ lines)} 
\ee
where 
coherent scattering occurs only by diffusion in space
and lines scatter by diffusion in space
and redistribution in 
frequency.  Wings of strong spectral lines form under conditions close to CS, the cores more like CRD - the $1/\epsilon$ factor applies to typical cases 
where ``partial redistribution'' is important \citep{Mihalas1978}. 
The dependence of $\tau_{th}$ on $\epsilon$ is stronger for
permitted line transitions instead of strict CS as
photons are redistributed in frequency as well as space.  As a
consequence, for lines, one must consider the frequency averaged mean
intensity $\overline J$ not just $J_\nu$.  For strong scattering
($\epsilon \ll 1$), the $\Lambda$ operator smears the source function
over a large number of mfps.  In the case of strong chromospheric
lines ({Ca~{\sc ii}} and Mg~{\sc ii} resonance lines, for example), the
thermalization depth spans most of the 1500 km thickness of the
stratified chromosphere \citep[{\em e.g.\rm}][]{Linsky+Ayres1978}.  In weaker
lines formed in the photosphere, such as spin forbidden transitions
({\em e.g.\rm} {Fe~{\sc i}} 524.7 and 525.0 nm),
$\epsilon $ is close to unity and the source function returns to LTE.

\noindent In 3D the microphysics is the same, but we must imagine instead a
`thermalization volume'' around a given point which contains all
sources that can influence the source function at that particular
point.  For a 3D stratified atmosphere 
this volume will look on average like a distorted egg with a squashed bottom and a stretched 
or open top (Figure~1).  The radiative transport becomes 
1D-like because lateral transport effects, in particular through the
$\Lambda$ operator, are less efficient than vertical escape of radiation.

\figthree

Figure~2 shows results of a Monte Carlo
calculation of the emergent intensity from the thermal sources shown
in Figure~1 for different values of the destruction probability 
$\epsilon$, assuming CS.    
The calculation has been set up to make any scattering effects
important, by placing the sources both beneath and above the optical depth unity
surface.    (If deeper, the mfps would be very small, if higher, the
mfps would be very large and the radiation optically thin).  


\noindent Differences between these calculations simply reflect the
scattering process {\em in the Sun itself}.  
Sub-mfp scales can be resolved, but enhanced scattering in the Sun
produces a diffuse image that appears qualitatively like a telescope
that has broad wings in the PSF (see Fig.~2), at least for features 
formed below the vertical $\tau=1$ surface.  The smoothing 
is on scales smaller than $\lambda \approx H$ in part because the 
outward directed
photons tend to escape and be observed with fewer scatterings than
inward or horizontally directed photons.  Therefore, the lateral
effects of scattering are smaller than in non-stratified cases.
This is a well known result used in early days to justify 
calculations with transport only in the vertical (radial) direction.
In Section \pref{ssec:mimick}
we discuss some scalings arising from these
calculations.

\subsection{Line transfer scalings}

The lines listed in Table~1 are all
electric dipole transitions of abundant elements.  
To estimate scattering parameters for those fully permitted lines
(i.e. excluding spin-forbidden lines of 524.7, 525.0 and 676.8 nm), 
we adopt the estimate of \cite{VanRegemorter1962}, in which the
long-range interaction between the incoming electrons and the atom
dominates the cross section, and produces collision probabilities that
are proportional to the radiative transition probability.  
Van Regemorter's \nocite{VanRegemorter1962} formula is
\be \label{eq:vr}
C_{ji} = 20.60 \lambda^3 T_e^{-1/2} p_z\left ({hc}/{\lambda kT }\right)\ A_{ji}
\ee
where $\lambda$ is in cm, $T_e$ in K, and 
$p_z(\beta)$ are weakly decreasing functions of $\beta$ for
ions (charge $z\ge 1$), and another that varies asymptotically as $\beta^{-1/2}$ for large 
$\beta$ in neutrals ($z=0$). 
The Maxwellian-averaged transition probability for collisional
de\--excitation by electrons of number density $n_e$ 
between the atomic levels $i$ and $j$ is ${\cal C}_{ji}=n_e{C_{ji}}$
s$^{-1}$, and the spontaneous radiative emission coefficient is
${A_{ji}}$ s$^{-1}$.
\figvr
Van Regemorter's equation (22) gives
\be
\epsilon^\prime = \frac{n_eC_{ji}}{A_{ji}} \, =\, 20.60 n_e \lambda^3 T_e^{-1/2} p_z\left ({hc}/{\lambda kT }\right),
\ee
for the ratio of collision de-excitation rates by electrons of number density $n_e$ to ${A_{ji}}$.   
$\epsilon^\prime$ depends only  weakly on $T_e$ which in any case
changes far less than the electron density across the solar
photosphere/chromosphere.  
The destruction probability  is 
\be \epsilon =
\frac{n_eC_{ji}}{{n_eC_{ji}}+A_{ji}} =
\frac{\epsilon^\prime}{1+\epsilon^\prime}. \ee
Estimates of $\epsilon$ for the Sun's atmosphere are shown in
Figure~3, using the stratified model 'C' of
\citet{Vernazza+Avrett+Loeser1981}.  Lines of neutrals differ from
those of charged ions because of the different behavior of cross
sections at threshold \citep{Seaton1962a}.  For the 
transitions involving changes of spin-- 524.7, 525.0 and 676.8 nm-- 
 we assumed cross sections of $\pi a_0^2$ with $a_0$
the Bohr radius, since such transitions require penetration of the target atom
by the electron (long range dipole interactions are small). 
The $\epsilon$ values are probably no better than
order of magnitude estimates given the approximations involved \citep{Allen1973}. Since
collisions with other particles and to other atomic levels, and
background continuum absorption are not included, they are probably
under-estimates by factors between one and two. 

Values of $\epsilon$ listed in Table~1 vary between 
one (spin-forbidden transitions) to $10^{-4.5}$ (strong lines with 
contributions to their formation  above heights of 500 km).   
From these calculations we see that, in order to minimize the 
kind of scattering effects shown in Figure~2, 
the spin-forbidden lines 
with $\epsilon$ parameters close to unity
should be used.  Indeed, differences between these 
the stronger permitted lines 
with far smaller values of $\epsilon$ should be exploited in
observational campaigns with high resolution telescopes. 

\subsection{3D Line transfer calculations in magnetoconvection simulations}

We have made calculations using the 422.7 nm
line of {Ca~{\sc i}} using the 3D nLTE radiative transfer program
``RH'' of \citet{Uitenbroek2000}, in magnetoconvection models
generated by using the MURaM code by one of us \citep{Rempel2014}. These are
small scale dynamo models ({\em i.e.\rm} a mixed polarity field is maintained
self-consistently by turbulent flows) in a $6.144 \times 6.144 \times
3.072\,\mbox{Mm}^3$ domain with a grid spacing of $8$~km. The radiation field 
is treated using a gray opacity, assuming LTE.   The average
$\tau=1$ surface is located about $700$~km beneath the top boundary.
These models include all thermal structure and
vector velocity and magnetic fields, but we have here
simply looked at the effects of radiation transport
ignoring magnetic and velocity fields, keeping only the 3D
thermal structure, for simplicity.  Both magnetic and velocity fields distribute 
opacity over broader wavelength ranges, thereby enhancing radiative transport. Our calculalations can
be regarded as minimizing the multi-dimensional transfer effects. 

{Ca~{\sc i}} was selected since it is a resonance line formed both in the 
photosphere and lower chromosphere.
We made some trade-offs with regard to the atomic model used for our
computations.  The calculations are CPU intensive using a large atomic
model, and so we included just the two levels of the {Ca~{\sc i}} 422.7 nm
(resonance) line and the {Ca~{\sc ii}} continuum state.  Another trade-off was
the use of a factor of 3 reduction in abundance (we used 5.86 on a scale
where H$\equiv12$), in order to reduce the
optical depth at the very top of the simulations. 
Our calculations are
therefore not to be taken as ``synthetic {Ca~{\sc i}}'' data but as numerical
experiments designed to examine the effects of the diffusion of photons in
space and frequency for conditions of interest in the Sun's
atmosphere.

\figthreed

Figure~4 compares the results from 1D and 3D calculations for several 
different values of the parameter $\epsilon$, at the center of
the {Ca \sc i} line.   The nominal value
of $\epsilon$ uses the collision rates in the atomic model 
included with RH- it amounts to a cross section about $3\times$ larger than
the Van Regemorter formula would give. 
Values of $\epsilon$ were changed simply by multiplying the 
collision rates between the two singlet levels by factors 
of $10^{2}$ and $10^{-2}$.

Panels (b) and (e) have the closest 1D vs. 3D 
properties, as expected since the line is formed closest to
LTE in both cases.   Nevertheless, the  3D 
images are still blurry compared with 1D. Notice that the 
lines are a factor of 3-4 brighter than the other four calculations 
that are farther from LTE.  
Panels (a) and (d) show remarkable qualitative differences
attributable to horizontal transfer, indeed on these scales one might
wonder if one were looking at the same region of the Sun.   (c) and (f),
furthest from LTE, are also radically different, with the 
$\epsilon=0.01$ 3D case being some 30\% dimmer than, with  less
contrast than the 3D case for $\epsilon=1$.  

The differences in images generated at different wavelengths 
(not shown) are progressively smaller 
as wavelengths move from the core to the wings. 
This is expected since the line opacity is relatively
smaller and the LTE continuum, dominated by H$^-$ absorption, making
significant contributions to both source function and optical depth.
Images made 
in 3D at 0.01 nm to the blue of line center are similar in overall
appearance to panel (c), panel (c) therefore shows structure a little deeper
in the atmosphere compared with the others shown.  At these inner wing
wavelengths,  qualitatively similar  effects are seen as in the line core
in terms of blurring, but the images are more similar to one another.

The images all show structure below the vertical 
photon mfp.  However, the  image contrasts are 
reduced compared with 1D calculations owing to horizontal 
radiative transfer and they can even have a very different morphology.  
The former results are consistent with
the narrative by \citet{Kneer1981}.   The effects of scattering cannot
therefore automatically be neglected, except perhaps for the continua
(H$^-$) and weak photospheric lines formed strictly in LTE.  They
become more important as stronger features form successively higher in
the atmosphere as the densities decrease with height. \citet{Leenaarts+others2010} indeed report ``halos'' around 
features in the very low $\epsilon$ {Na~{\sc i}} D lines in their radiation MHD
calculations.  

\section{Discussion}
\label{sec:discussion}

\subsection{nLTE transfer can mimick a poor telescope PSF}
\label{ssec:mimick}

The scattering intrinsic to solar plasma 
has a similar effect to reducing the imaging performance of a
telescope.  This was explored by \citet{Kneer1981} in terms of a
modulation transfer function $M_I(k)$ (MTF):
\begin{equation}
M_I(k)= I'(k)/I'(k=0)
\end{equation}
where 
the Planck function is assumed to vary horizontally harmonically as 
$a +b~\cos~kx$ ($x$=horizontal direction),
$I' = I - \langle I\rangle$, $I$ the output intensity and
$\langle I \rangle$ the horizontally averaged output intensity.  
His figures 2 and 5 compare the MTFs 
of CS and CRD calculations with a 1m telescope MTF, showing the dramatic
drop of the MTFs with decreasing $\epsilon$ and increasing $k$ values, and the smaller MTFs
for CRD versus CS. 
The Hankel transform of
the MTF gives the point spread function (PSF) of the scattering
atmosphere.  His figures 3 and 6 compare the computed PSFs  using values of $\epsilon$ from $10^{-6}$ to
$10^{-1}$, with telescope PSFs for various length scales $k^{-1}$.
The figures show the detrimental effects of the
scattering clearly broadens the PSF as $\epsilon$ decreases.  Also, his
Figure~7 shows the computed full width at half maximum (FWHM) of the
PSFs as functions of $\tau_{th}$ which have the dependence of
equation~\pref{eq:scaledep}.  

For our
purposes we use the simple definition of the Strehl
ratio as the ratio of the peak image
intensity from a point source to the maximum attainable
intensity using an ideal optical system. We then estimate for each
synthetic calculation (independent of any telescope system) 
an equivalent Strehl ratio. We use intensities along vertical rays. 

\figscales

Figure~6 shows that 
the equivalent Strehl ratio $\cal S$  for CS for the embedded sources used 
varies for large $\epsilon$ as 
$$
{\cal S} \approx \sqrt \epsilon^{}, \ \ \ \ {\rm for\ \epsilon > 0.1}
$$
For smaller
values of $\epsilon$, ${\cal S}$ is weakly dependent on $\epsilon$, being 
$> 0.1$ for all solar lines of interest.  
Figure~2 shows that core intensity is moved laterally by up to 1
(vertical) mfp from its thermal source for almost all values of
$\epsilon$. Taken together, we see that sub$\lambda$ scale, high contrast sources
should be distinguishable even in the presence of scattering.  Kneer's
Figures 4 and 7 in fact already exhibit a similar result (although
Kneer himself does not report this).  In his case, the central values
of the PSFs shown seem to scale with $\epsilon^{\alpha}$ with 
$\alpha \approx 0.3-0.4$, as shown in our Figure 6. (Kneer does not show results for $\epsilon=1$,
therefore 
we have moved his two curves upwards by 0.2 dex so that his CS calculations agree with ours at 
$\epsilon=0.1$).  The differences between the calculations are real. 
We would not expect the curves to agree - these are very different 
calculations (Kneer has periodic sources spread through his domain, we have point sources 
at special places near optical depth unity). But the trends are clear, the Strehl ratio decreases
initially like $\epsilon^\alpha$ before beginning to flatten at very small $\epsilon$ values. 
We believe flattening arises from stratification which, no
matter the number of scatterings, tends to promote photon escape vertically out
of the atmosphere. 

\subsection{A simple conceptual picture}

With future telescopes we will routinely observe the Sun's atmosphere at 
scales $\ll \lambda$.  In our 2D calculations we have considered isolated
bright points  separated by length scales 
$\delta < \lambda$.  When we have a thermal source
function, even when $\delta \gg \lambda$ the two structures are fully
resolved since either the rays intercept the hot structure or they do
not, and we assume the source function is negligible outside of the
two points.  Even if the source function outside is finite, 
points will be resolved if their brightness is sufficient to
discriminate them from each other and the background.  The resolution
along one single line of sight remains about one pressure scale height
without stereoscopy \citep{Landi2013}.

The situation is different in the presence of scattering, since the
hot source function ``leaks'' into neighboring regions of the
atmosphere with, in 1D, a characteristic length scale for
thermalization of $\lambda/\epsilon^\alpha$.  Our 2D CS  
calculations suggest that the horizontal scattering should
spread the light finally emerging from point sources by about 1
vertical mfp from their origin, and that their ``equivalent Strehl
ratio'' should be around ${\epsilon}^{\alpha}$,
$\alpha=1/3$ to $1/2$. Using Kneer's earlier results 
the Strehl ratio would be, for CRD in lines, about $\epsilon^{0.44}$.

In 3D the situation is
more complex and we can estimate that the characteristic length will
be some number $x$ times $\lambda$, $x \gtrsim 1$.  If $\delta
\lesssim x\lambda$ then some loss of resolution will result.
Depending on the 3D thermal structure and scattering parameters
though, it may still be possible to resolve the two bright sources,
like being able to see two car headlights through a dense fog.  Our 3D
calculations for a  simple {Ca~{\sc i}} atom support these conclusions.   
The resolution along one single line of sight in this case is reduced
to $\lambda/\epsilon^\alpha$, but since the mfps exponentially
decrease with depth, this means the LOS resolution may not be that
dissimilar from the thermal case.

\subsection{Some outstanding questions}

There are cases that might appear to contradict the
above discussion.  These cases seem to require decoupling of optical
depth scale from the source function in the optically thin parts of an
atmosphere, allowing some fine structures to be imprinted as
absorption (or emission) features on a
smoother, spatially smeared scattering source function.

Consider the observation of  
a terrestrial cloud passing between the Sun and telescope. 
We can resolve the cloud structure down to
the 
telescope's resolution limit, without regard to the intensity
$I_\odot$ of the background illuminating 
source (the Sun).   In this case the source function in the cloud is
negligible, the cloud's signature is made
clear in the image only through the term
$$
I = I_\odot e^{-\tau_{cloud}}
$$
Thus the measured intensity can have any scale the telescope can
resolve along different lines of sight, independent of $I_\odot$.   

Next imagine an an optically thin ``cloud'' (spicule, filament, fibril) above
the Sun's photosphere.  We might expect the above terrestrial argument to hold. But
it must be remembered that close to the Sun there is a low
temperature limit to plasma sitting above the solar surface.  In the
absence of strong adiabatic cooling due to dynamics, the source
function at these heights can approach the radiative equilibrium
temperature which, for an atmosphere with gray opacity, is near 4000K.  The above equation must include this
emission.  Let the optical depth of a ``spicule'' be $\tau_s < 1$  
so that with $I_\odot \approx B(5800K)$,  
$$
I = B(5800K) e^{-\tau_s} + B(4000K) \tau_s
$$
In this case the contrast of the solar ``cloud'' will be reduced by the second
term but in principle, if very
fine structure absorbing clouds exist on the solar surface, then 
they will be visible against the solar disk.   This is indeed observed 
\citep[{\em e.g.\rm}][]{Sekse+others2012,Lipartito+others2014}. 

As a third example, we imagine observing strong lines (chromospheric
lines) far above the solar limb.  Source functions are determined by
local conditions {\em and} the radiation within the emitting/absorbing
atmospheric structure itself
(this is also the case for the disk atmosphere, it's just that here we
imagine observing this plasma mostly tangentially not radially to the
solar surface where the background intensity is zero).  The opacities are also determined in this fashion,
but there are classes of strong spectral lines for which the ambient
radiation field is less important in the opacity than the source
function.  Consider the $h$ and $k$ lines of Mg~{\sc ii} for example.  The
lower level populations of Mg~{\sc ii} throughout the chromosphere are
essentially decoupled from the source functions of these lines
themselves, being simply proportional to the number density of Mg$^+$.
This number density can be determined by local conditions.
In this situation, very fine structure in ``threads'' or
``blobs'' will be resolved by the telescope independent of the mfp of
photons, since the intensity is either 
$$
I = \int S d\tau_{LOS} \ \ \ {\rm \ optically\ thin}
$$
or 
$$
I \approx  S(\tau_{LOS}=1)   \ \ \ {\rm \ optically\ thick}
$$
Again,
either a ray intercepts some plasma - influencing $\tau$ and $S$
(which may itself be smeared), or it does not.  
Thus if the differences in $\tau$ between rays are large then
in both thick and thin cases the telescope should resolve these
features.  

This same optically thin argument applies to a fourth case, 
optically thin coronal lines, where we
have seen structures down to $0.3''$ scales when the photon mfps
are almost infinite.  

A fifth kind of example involves decoupling of source functions and 
optical depths via Doppler shifts. 
In the case of spectral lines, the source function has not
just a $\Lambda$ operator smoothing variations in $S$ over depth, but
also it has integral operators coupling different frequencies.   We simply point out
that the local mfp varies greatly in frequency so that relatively 
modest Doppler shifts can decouple the optical depth scale from the 
source function.  Under such circumstances this decoupling 
can make the optical depths between adjacent rays through plasma 
moving with
different Doppler velocities very different,
analogous to the cloud cases discussed above but for a slightly
different reason.  Again such structure is observed 
\citep{Judge+Reardon+Cauzzi2012}. 

These examples reveal that one must carefully understand the nature of
the source functions when observing below typical photon mean free paths.

\section{Conclusions}\label{s:Conclusions} 

The Sun itself sets fundamental limitations on the ultimate
``resolution'' that can be obtained, even in the photosphere where LTE
is often assumed.   On scales below photon mean free paths, in the
presence of scattering in typical permitted photospheric lines, LTE
breaks down since the source function appears to be in LTE only on
larger physical scales where $J$ can approach $B$.  The decoupling
of source function from local conditions produces smeared 
images which are qualitatively similar to those produced with a
telescope PSF which has broad wings.   We have 
tried to unify the literature by drawing attention to the different scattering parameters
used by various workers and by using MTF and PSF characterizations of 
the atmosphere, following \citep{Kneer1981}.  We have identified some simple scaling laws that
might be useful when observing at and below the pressure scale height. 
On the basis of our study, we suggest that simultaneous imaging with 
both strongly and weakly scattering lines be conducted with 
high resolution telescopes.   These might include 
several of the permitted {Fe \sc i} lines and the two 
spin-forbidden {Fe \sc i} lines at 524.7 and 525.0 nm.  

The effects discussed are non-LTE effects of one particular
kind, represented by
equation~(\pref{eq:parameter}). These are nLTE effects entering 
through the 
{\em source
  function}.   Other multi-level effects include   
over-ionization of neutral species by non-LTE ionizing radiation,
affecting primarily the {\em optical depth scale}, and the 
(implicit) influences of other bound-bound radiative transitions on the 
source function of a particular line.  In principle, such effects should add to the smearing
effects outlined here through the additional non-local processes involved.

%
 \begin{acks}
Philip Judge is very grateful to NAOJ for support of a Visiting 
Professorship there during July and August of 2012, where this
work was begun.  
 \end{acks}

%
\appendix   

\section{Mean free paths and scale heights}
\label{sec:emerging}

Photons emerging from an optically thick atmosphere arise mostly from regions
centered around optical depth $\tau=1$.  As 
\cite{Landi2013} has pointed out, in stratified
layers this has the curious property that {\em irrespective of the
  strength of the transition}, the photons emerge from a region with
an intrinsic thickness of $\approx 1$ pressure scale height, when the opacity
is proportional to pressure.   
This scale represents a
basic limit below which one cannot say with certainty that a certain
photon arose from or ``forms at'' a specific height.  This statement applies to all
features formed in optically thick stratified layers, in particular 
throughout the photosphere and the subsonic
(non-spicular) components of the chromosphere\footnote{See appendix B.}.

We can show this result formally.  Let $z$ measure local height above some
arbitrary point in the Sun, with the observer at $\infty$.  The
photosphere and low chromosphere comprise
a partially ionized stratified layer with $T\approx 6000$ K,
and a pressure scale height
\be H = - \left (\frac{dln~p}{dz} \right
)^{-1} = \frac{kT}{\mu m_H g} \approx 130 {\rm \ \ \ km}. \ee 
Across the solar photosphere-chromosphere, temperature changes by less
than a factor of two whereas the pressure changes by orders of
magnitude. Thus both gas density and pressure vary roughly as
$e^{-z/H}$.
Define the
optical depth as usual, 
\be \label{eq:tau}
\tau(z) = - \int_\infty^z k(s) ds
\ee
\noindent where $k(s) = \lambda(s) ^{-1}$ and $\lambda(s)$ is the
local mean free path of photons.  We can also
factor the opacity into $k(s) = \kappa(s) \rho(s)$ where $\kappa$ is
the opacity in cm$^2$g$^{-1}$, often a slowly varying quantity, and
$\rho(s)$ the mass density in g~cm$^{-3}$.

Define the region
from which most photons emerge as between, say, $\tau=1/{\rm e}$ and
$\tau={\rm e}$. 
The geometrical thickness of this region is  $\Delta=z_1-z_2$
where
\be  \label{eq:width}
{\rm e}^{-1} = -\int_\infty^{z_1} k(s) ds,  \ \ \ \ 
{\rm e}^{+1} = -\int_\infty^{z_2} k(s) ds.
\ee
In the photosphere, opacity is dominated by the minority species H$^-$
\citep[{\em e.g.\rm}][]{Allen1973}.
In this case $\kappa \propto n_e$ where $n_e$ is the electron
density, and in this case, $n_e \propto \rho$ 
since electrons come from singly ionized Fe, Si\ldots.  (There is also 
a temperature dependence through the Saha H, $e$ and H$^{-}$ equilibrium
which is important in deep photospheric layers, ignored here for simplicity). 
Above the
photosphere, $\kappa$  is
dominated by lines and continua of neutrals and singly ionized ions
($H^-$ being negligible at smaller densities), and the opacity
$\kappa(s) \approx constant$ \citep[{\em e.g.\rm}][]{Vernazza+Avrett+Loeser1981}.  
We can write, roughly, 
$$
\kappa(s) = \kappa(0) \left ( \frac{\rho(s)}{\rho(0)} \right ) ^m
$$
where $m=1$ (photosphere: H$^-$) and $m=0$ (chromosphere).  
So for a simply stratified atmosphere we have 
\be \label{eq:ks}
k(s) = k(0) {\rm e}^{-(m+1)s/H}.
\ee
Combining with eq.~(\pref{eq:width}) we have 
\be 
{\rm e}^{-1} = \frac{k(0) H}{m+1} e^{-z_1(m+1)/H},  \ \ \ \ 
{\rm e}^{+1} = \frac{k(0) H}{m+1} e^{-z_2(m+1)/H}.  \ \ \ \ 
\ee
The thickness of the region emitting most observed photons ({\em i.e.\rm} those actually
emerging from the atmosphere) is
simply
\be
\Delta =(z_1-z_2) =  2H/(m+1).
\ee
Now, the mean free path of a photon where $\tau=1$ is $1/k(\tau=1)$ 
must be $\approx \Delta$:  if it were larger (say $10\Delta$) then this
would mean that photons 10x deeper would escape, contrary to our
assertion that they escape from $\tau=1$; if smaller, the photons
would not escape, thus
\be
\lambda \approx \Delta = 2H/(m+1).
\ee
(The factor of two arises simply because of our adoption of brightness 
factors ${\rm e}$ and ${\rm 1/e}$).   
The photon path length is of the same order as the
pressure and density scale height in the photosphere ($m=1$)
and lower chromosphere ($m=0$).  

\section{Surely the chromosphere is not hydrostatic?}

\noindent When fluid motions are subsonic, the atmosphere cannot be far
from a hydrostatic state.  This is because sound/ magnetosonic waves
propagate changes in pressure to make the atmosphere approach a balance
between pressure gradients and gravity.  Much attention has been
drawn to chromospheric features with highly 
supersonic motions.  The classic
semi-empirical, hydrostatic models of VAL
\citep[{\em e.g.\rm}][]{Vernazza+Avrett+Loeser1981} 
 have been much maligned, the chromosphere
at the limb appears at a first glance to be extended much beyond the
1.5 Mm thickness of the model.  So what is going on here?

Those features moving supersonically include spicules and
rapidly moving features seen on the solar disk.  We can estimate what
fraction of the Sun is covered at any given time by these rapid events
using numbers recently published 
by Sekse and colleagues 
\citep{Sekse+others2012}, revising earlier estimates 
by \citet{Judge+Carlsson2010}.
The former authors find up to $2\times10^5$ of
these
features on
the Sun at any given time.  Each is typically $\ell=$2.5 Mm long and say at
most $w=0.2$ Mm across.  Thus the area covered by these features is $< w\ell = 10^5$
Mm$^2$, of a total solar surface area of $6\times10^6$ Mm$^2$.  Thus
by these data, 
at most just $1.7\%$ of the Sun's chromosphere has supersonic motions
associated with it.  This appears broadly consistent with the images
shown by these authors, only a small fraction of the solar surface 
shows these features. 

However it has been claimed by some that there is ``no
chromosphere'' when there is no clear fibril structure. Given that
many fibrils are supersonic, the claim might be extended to mean that
there is no hydrostatic layer.  It is difficult to assess what this
claim might mean since the Sun cannot go immediately from photospheric
pressures to coronal pressures with no intervening plasma pressure or
magnetic stress- they differ by five (!) orders of magnitude.
While claims and opinions may change, data do not. 
It is clear, and has been for some time, that if one observes narrow
chromospheric line profiles on the solar disk, features with
supersonic motions are quite rare. Using observations of C~{\sc i} emission
emission lines from HRTS, for example, \citet{Dere+others1983}
found that supersonic chromospheric ``jets'' are born at a rate of
some $3\times10^{-4}$ Mm$^{-2}$~s$^{-1}$ with a lifetime of 40s or
so, and each has an area of some 1 Mm$^2$.  Using these figures we
see that they cover just $1\%$ or so of the total area of the
chromosphere.  

Therefore, we must recognize that the bulk of
the chromosphere is approximately stratified in hydrostatic equilibrium. 
This is also consistent with early analyses of the solar flash
spectrum \citep{Athay1976} and the existence of 3-5 min oscillations throughout much of
the chromosphere requires a hydrostatic stratification.   
As far as the chromosphere is concerned, the supersonic features are
interesting but fill only a tiny fraction of the entire chromosphere.

\nocite{Judge+Carlsson2010}
\nocite{Lipartito+others2014}

%
%
%

\end{article} 
\end{document}